\documentclass[a4paper]{llncs}
\usepackage{graphicx}
\usepackage{caption}
\usepackage{url}
\usepackage{subcaption}
\usepackage{subfig}

\begin{document}
\mainmatter

\title{BiDAl: Big Data Analyzer for Cluster Traces\thanks{This is an
    author-generated version of a paper published and copyrighted by
    Gesellschaft f\"ur Informatik e.V. (GI). The copyright holder
    grants the authors the right to republish this work. Please cite
    as: Alkida Balliu, Dennis Olivetti, Ozalp Babaoglu, Moreno
    Marzolla, Alina S\^irbu, BiDAl: Big Data Analyzer for Cluster
    Traces, in E. Pl\"odereder, L. Grunske, E. Schneider, D. Ull
    (editors), proc. \emph{INFORMATIK 2014 Workshop on System Software
      Support for Big Data (BigSys 2014)}, September 25--26 2014,
    Stuttgart, Germany, Lecture Notes in Informatics (LNI) --
    Proceedings, Series of the Gesellschaft f\"ur Informatik (GI),
    Volume P-232, pp. 1781--1795, ISBN 978-3-88579-626-8, ISSN
    1617-5468}}

\author{Alkida Balliu, Dennis Olivetti, Ozalp Babaoglu, Moreno Marzolla, Alina S\^{i}rbu}

\institute{Department of Computer Science and Engineering, University of Bologna\\
  Mura Anteo Zamboni 7,
  40126 Bologna, Italy \\
  \{alkida.balliu, dennis.olivetti\}@studio.unibo.it \\
  \{ozalp.babaoglu, moreno.marzolla, alina.sirbu\}@unibo.it
}

\maketitle

\begin{abstract}
Modern data centers that provide Internet-scale services are stadium-size structures housing tens of thousands of heterogeneous devices (server clusters, networking equipment, power and cooling infrastructures) that must operate continuously and reliably.  As part of their operation, these devices produce large amounts of data in the form of event and error logs that are essential not only for identifying problems but also for improving data center efficiency and management. These activities employ data analytics and often exploit hidden statistical patterns and correlations among different factors present in the data.  Uncovering these patterns and correlations is challenging due to the sheer volume of data to be analyzed. This paper presents BiDAl, a prototype ``log-data analysis framework'' that incorporates various Big Data technologies to simplify the analysis of data traces from large clusters. BiDAl is written in Java with a modular and extensible architecture so that different storage backends (currently, HDFS and SQLite are supported), as well as different analysis languages (current implementation supports SQL, R and Hadoop MapReduce) can be easily selected as appropriate. We present the design of BiDAl and describe our experience using it to analyze several public traces of Google data clusters for building a simulation model capable of reproducing observed behavior.
\end{abstract}

\section{Introduction}

Modern Internet-based services such as cloud computing, social networks, online storage, media-sharing, etc., produce enormous amounts of data, not only in terms of user-generated content, but also in the form of usage activity and error logs produced by the devices implementing them. Data centers providing these services contain tens of thousands of computers and other components (e.g., networking equipment, power distribution, air conditioning) that may interact in subtle and unintended ways, making management of the global infrastructure far from straightforward. At the same time, services provided by these huge infrastructures have become vital not only to industry but to society in general, making failures extremely costly both for data center operators and their customers.  In this light, monitoring and administering data centers become critical tasks. Some aspects of management, like job scheduling, can be highly automated while others, such as recovery from failures, remain highly dependent on human intervention. The ``holy grail'' of system management is to render data centers autonomous, self-managing and self-healing; ideally, the system should be capable of analyzing its state and use this information to identify performance or reliability problems and correct them or alert system managers directing them to the root causes of the problem. Even better, the system should be capable of anticipating situations that may lead to performance problems or failures, allowing for proactive countermeasures to steer the system back towards desired operational states. Needless to say, these are very challenging goals \cite{Salfner2010}.

Given the size of modern data centers, the amount of log data they produce is growing steadily, making log management itself technically challenging. For instance, a 2010 Facebook study reports 60 Terabytes of log data being produced by its data centers each day~\cite{Thusoo2010}.  For live monitoring of its systems and analyzing their log data, Facebook has developed a dedicated software called Scuba~\cite{Abraham2013} that uses a large in-memory database running on hundreds of servers with 144~GB of RAM each. This infrastructure needs to be upgraded every few weeks to keep up with the increasing computational power and storage requirements that Scuba generates.  Log analysis falls within the class of Big Data applications: the data sets are so large that conventional storage and analysis techniques are not appropriate to process them.  There is a real need to develop novel tools and techniques for analyzing logs, possibly incorporating data analytics to uncover hidden patterns and correlations that can help system administrators avoid critical states, or to identify the root cause of failures or performance problems.

Numerous studies have analyzed trace data from a variety of sources for different purposes, but typically without relying on an integrated software framework developed specifically for log analysis~\cite{Chen2012,Liu2012,Reiss2012}.  This is partially due to the sensitive nature of commercial log trace data prohibiting their publication, which in turn leads to fragmentation of analysis frameworks and difficulty in porting them to traces from other sources.  One isolated example of an analysis framework is the Failure Trace Archive Toolkit~\cite{Javadi2013}, limited however to failure traces. Lack of a more general framework for log data analysis results in time being wasted by researchers in developing software for parsing, interpreting and analysing the data, repeatedly for each new trace~\cite{Javadi2013}. 

In this paper we describe the Big Data Analyzer (BiDAl), a prototype software tool implementing a general framework, designed for statistical analysis of very large trace data sets. BiDAl integrates several built-in storage types and processing frameworks and can be easily extended to support others. The BiDAl prototype is  publicly available through a GNU General Public License (GPL)~\cite{bidalCode}. We illustrate the actual use of BiDAl for analyzing publicly-available Google cluster trace data~\cite{googleData} in order to extract parameters for a cluster simulator which we have implemented.

The contributions of this work are several fold.  We first present the novel architecture of BiDAl resulting in extensibility and ease of use. BiDAl incorporates several advanced Big Data technologies to facilitate efficient processing of large datasets for data analytics. We then describe our experience with BiDAl when used to extract workload parameters from Google compute cluster traces. Finally, we describe a simulation model of the Google cluster that, when instantiated with the parameters obtained through BiDAl, is able to reproduce a set of behaviors very similar to those observed in the traces.

The rest of the paper is organized as follows.  We provide a high level overview of the framework followed by a detailed description of its components in Section~\ref{bidal}. The framework is applied to characterize machines and workloads in a public Google cluster trace, and used in the development of a cluster simulator in Section~\ref{testcase}. We discuss related work in Section~\ref{refs} and conclude with new directions for future work in Section~\ref{conclusions}.

\section{The Big Data Analyzer (BiDAl) prototype}\label{bidal}

\subsection{General overview}

BiDAl can import raw data in CSV format (Comma Separated Values, the typical format of trace data), and store it in different backends according to the user's preference.  In the current prototype two backends are supported: SQLite and Hadoop File System (HDFS), the latter being particularly well suited for handling large amount of data using the Hadoop framework. Other backends can easily be supported, since BiDAl is based on a modular architecture that will be described in the next section. BiDAl uses a subset of the SQL language to handle the data (e.g., to create new tables or to apply simple filters to existing data). SQL queries are automatically translated into the query language supported by the underlying storage system (RSQLite or RHadoop).  

BiDAl also has the ability to perform statistical data analysis using both R~\cite{r} and Hadoop MapReduce~\cite{hadoop,mapreduce} commands.  R commands are typically applied to the SQLite storage, while MapReduce to the Hadoop storage. However, the system allows mixed execution of both types of commands regardless of the storage used, being able to switch between backends (by exporting data) transparently to the user. For instance, after a MapReduce operation, it is possible to analyze the outcome using R; in this case, the software automatically exports the result obtained from the MapReduce step, and imports it to the SQLite storage where the analysis can continue using R commands. This is particularly useful for handling large datasets, since the volume of data can be reduced by applying a first processing step with Hadoop/MapReduce, and then using R to complete the analysis on the resulting (smaller) dataset.

\subsection{Design}

\begin{figure}[htb]
  \begin{center}
    \includegraphics[width=0.7\textwidth]{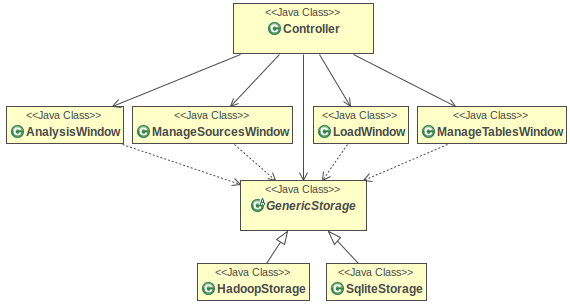}
    \caption{\label{uml}UML diagram of BiDAl classes.}
  \end{center}
\end{figure}

BiDAl is a modular application designed for extensibility and ease of use. It is written in Java, to facilitate portability across different Operating Systems, and uses a Graphical User Interface (GUI) based on the standard Model View Controller (MVC) architectural pattern. The View provides a Swing GUI, the Model manages different types of storage backends, and the Controller handles the interaction between the two. Figure~\ref{uml} outlines the architecture using the UML class diagram. 

The Controller class connects the GUI with the other components of the software. The Controller implements the Singleton pattern, with the one instance accessible from any part of the code. The interface to the different storage backends is given by the GenericStorage class, that has to be further specialized by any concrete backend developed. In our case, the two existing concrete storage backends are represented by the SqliteStorage class to support SQLite, and the HadoopStorage class, to support HDFS. Neither the Controller nor the GUI elements communicate directly with the concrete storage backends, but only with the abstract class GenericStorage. This simplifies the implementation of new backends without the need to change the Controller or GUI implementations.

The user can inspect and modify the data storage using a subset of SQL; the SqliteStorage and HadoopStorage classes use the open source SQL parser Akiban to convert the queries inserted by users into SQL trees that are further mapped to the native language (RSQLite or RHadoop) using the Visitor pattern. The HadoopStorage uses also a Bashexecuter that allows to load files on the HDFS using bash shell commands. A new storage class can be implemented by providing a suitable specialization of the GenericStorage class, including the mapping of the SQL tree to specific commands understood by the backend.
Although the SQL parser supports the full SQL language, the developer must define a mapping of the SQL tree into the language supported by the underlying storage; this often limits the number of SQL statements that can be supported due to the difficulty of realizing such a mapping.

\subsection{Functionality}
The typical BiDAl workflow consists of three steps: instantiation of a storage backend (or opening an existing one), data selection and aggregation and data analysis. For storage creation, BiDAl is designed to import CSV files into an SQLite database or to a  HDFS file system, depending on the type selected. Except for the CSV format, no other restrictions on the data type exist, so the platform can be easily used for data from various sources, as long as they can be viewed as CSV tables. Even though the storages currently implemented are based on the the concept of tables (stored in a relational database by SQLite and CSV files by Hadoop), in the future, other storage types can be supported by BiDAl. Indeed, Hadoop supports HBase, a non-relational database that works with $<$key, value$>$ pairs. Since Hadoop is already supported by BiDAl, a new storage that works on this type of non-relational databases can be easily added.

\begin{figure}[htb]
  \begin{center}
    \includegraphics[width=0.95\textwidth]{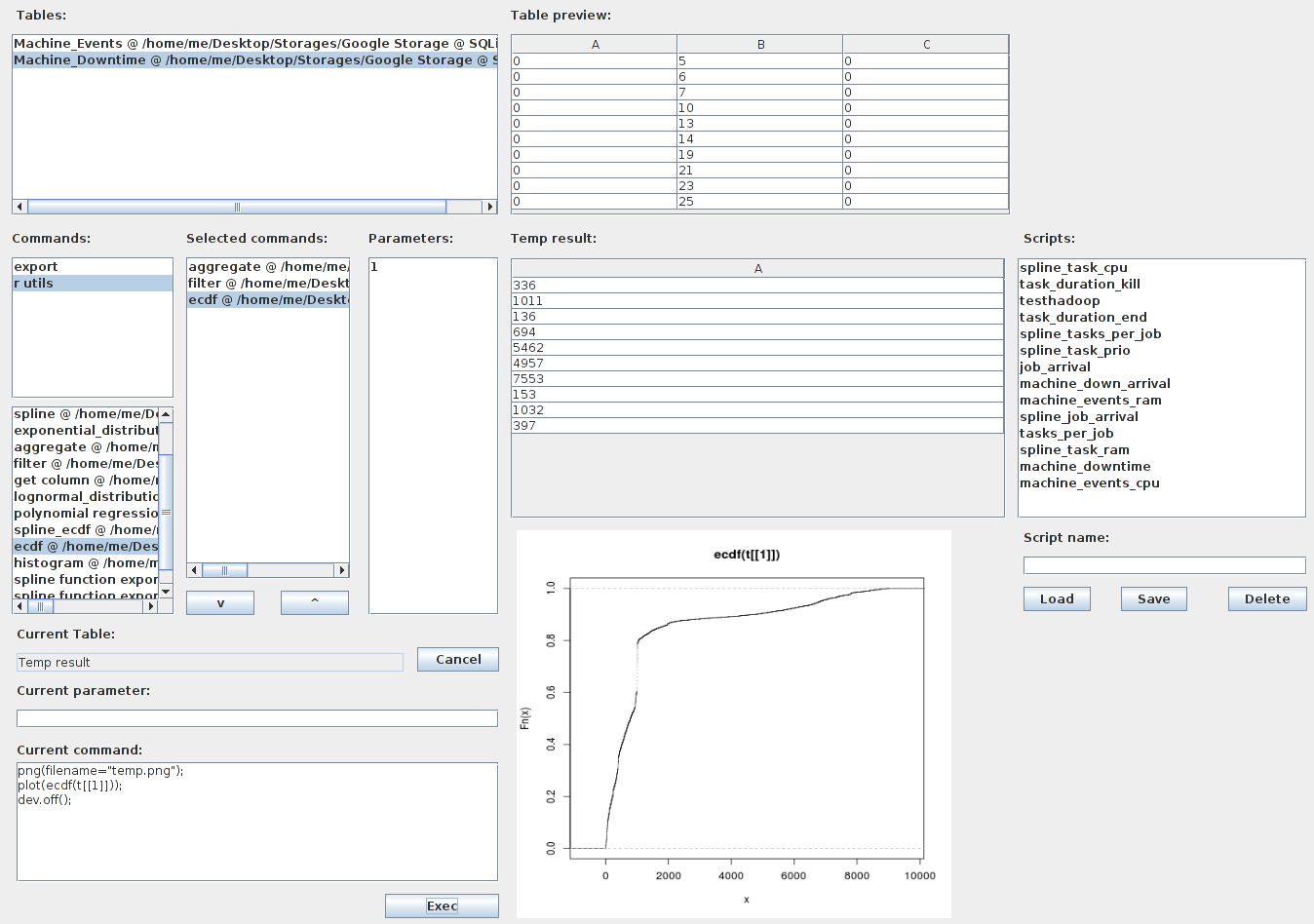}
    \caption{\label{gui}Screenshot of the BiDAl analysis console displaying R commands.}
  \end{center}
\end{figure}

Selection and aggregation can be performed using queries expressed using a subset of SQL. At the moment, the supported statements are SELECT, FROM, WHERE and GROUP BY. For the SQLite storage, queries are executed through the RSQLite library of the R package (R is used quite extensively inside BiDAl, and executing SQLite queries through R simplified the internal structure of BiDAl as we could reuse some internal software components). For the Hadoop backend, GROUP BY queries are mapped to MapReduce operations. The Map function implements the GROUP BY part of the query, while the Reduce function deals with the WHERE and SELECT clauses. We used RHadoop as a wrapper so that we can access the Hadoop framework through R commands. This allows the implementation of  Map and Reduce functions in R rather than Java code.

Data analysis can be performed by selecting different commands in the specific language of the storage and applying them to the selected dataset.  There is a common set of operations provided by every storage. However it is possible to concatenate operations on different storage backends  since BiDAl can automatically export data from one backend and import it on another. Therefore it is possible to use a MapReduce function on an SQLite table, or execute a R command on a HDFS store. This requires that the same data is duplicated into different storage types so, depending on the size of the dataset, additional storage space will be consumed. However, this operation does not generate consistency issues, since log data does not change once it is recorded.

\paragraph{Using R within BiDAl}
BiDAl provides a list of pre-defined operations, implemented in R, that can be selected by the user from a graphical interface (see Figure~\ref{gui} for a screenshot and Table~\ref{r} for a full list of available commands). When an operation is selected, an input box appears asking the user to provide the parameters needed by that specific operation. Additionally, a text box (bottom left of Figure~\ref{gui}) allows the user to modify on the fly the R commands to be executed.

\begin{table}
\centering
\small
\begin{tabular}{p{4cm}|p{8cm}}
\hline
\bf{R command} & \bf{Description}\\ \hline
get\_column&
Selects a column.\\ \hline
apply\_1Col&
Applies the desired function to each element of a column.\\ \hline
aggregate&
Takes as input a column to group by; among all rows selects the ones that satisfies the specified condition; the result obtained is specified from the function given to the third parameter.\\ \hline
difference\_between\_ rows&
Calculates the differences between consecutive rows.\\ \hline
filter&
Filters the data after the specified condition.\\ \hline
exponential\_distribution&
Plots the fit of the exponential distribution to the data.\\ \hline
lognormal\_distribution&
Plots the fit of the lognormal distribution to the data.\\ \hline
polynomial\_regression&
Plots the fit of the n-grade polynomial regression to the data in the specified column.\\ \hline
ecdf&
Plots the cumulative distribution function of the data in the specified column.\\ \hline
spline&
Divides the data in the specified column in n intervals and for each range plots spline functions. Also allows to show a part of the plot or all of it.\\ \hline

\end{tabular}
\caption{List of some R commands implemented by BiDAl.}\label{r}
\end{table}

All operations are defined in an external text file, according to the following BNF grammar:

\begin{footnotesize}
\begin{verbatim}<file> ::= <command name> <newline> <number of parameters> <newline>
       	   <list of parameters> <newline> <command code>
      	 
<list of parameters> ::=  <parameter description> <newline>
    			<list of parameters> | <empty>
                      	 
<command code> ::= <text> | <command code> <parameter>
	 		<command code> | <empty>
                	 
<parameter> ::= '$PAR' <number of the parameter> '$' 
\end{verbatim}
\end{footnotesize}

New operations can therefore be added quite easily by simply adding them to the file.

\paragraph{Using Hadoop/MapReduce with BiDAl}
BiDAl provides also a list of Hadoop/MapReduce commands that allow to distribute computation across several machines. Usually, the Mapper and Reducer functions are implemented in Java, generating files that need to be compiled and then executed. However, BiDAl abstracts from this approach by using the RHadoop library which handles MapReduce job submission and permits to interact with Hadoop's file system HDFS using R functions. Once the dataset of interest has been chosen, the user can execute the Map and Reduce functions already implemented in RHadoop or create new ones. Again, the MapReduce functions are saved in an external text files, using the same format described above, so the creation of new commands does not require any modification of BiDAl. At the moment, one Map function is implemented in BiDAl, which groups the data by the values of a column. The Reduce function counts the elements of each group. Other functions can be added by the user, similar to R commands. 

\section{Case study}\label{testcase}
The development of BiDAl was initially motivated by the need to process large data traces of compute clusters, such as those publicly released by Google \cite{googleData}. The ultimate goal was to extract workload parameters from the traces in order to instantiate a simulation model of the compute cluster capable of reproducing the most important features observed in the real data. The simulation model, then, can be used to perform ``what-if analyses'' by exploring different scenarios where the workload parameters are different, or several types of faults are injected into the system.
In this section we first describe the use of BiDAl for analyzing the Google traces, and then present the structure of the simulation model instantiated with the parameters obtained from the analysis phase.

\subsection{Workload Characterization of the Google Cluster}\label{workload}

\begin{table}
\centering
\small
\begin{tabular}{p{3cm}|p{6cm}|p{3cm}}
\hline
\bf{R command} & \bf{Parameter type}& \bf{Parameter value}\\ \hline
get\_column& column number&2\\ \hline
filter&condition& t[[1]]$<$11000.\\ \hline
log\_histogram&column number, log step, log axis& 1, 0.06, xy\\ \hline
\end{tabular}
\caption{Commands used to generate Figure~\ref{task}.}\label{ex}
\end{table}

\begin{figure}
  \begin{center}
     \begin{subfigure}{0.49\textwidth}
    \includegraphics[width=\textwidth]{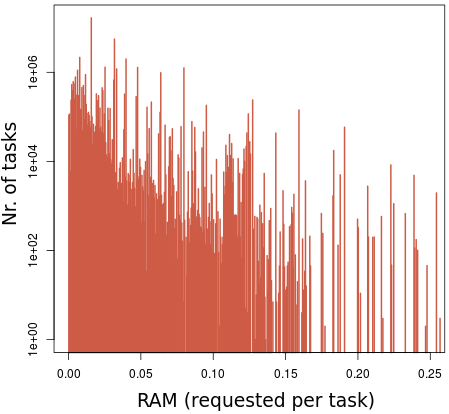}\caption{RAM requested by tasks. Values are normalized by the maximum RAM available on a single node in the Google cluster.}\label{ram}
    \end{subfigure}
     \begin{subfigure}{0.49\textwidth}
    \includegraphics[width=\textwidth]{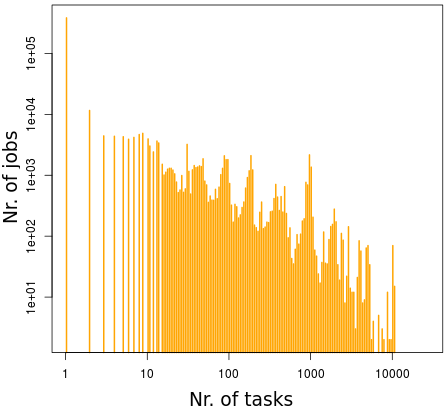}\caption{Number of tasks per job}\label{task}
    \end{subfigure}
    \caption{Examples of distributions obtained with BiDAl.}
  \end{center}
\end{figure}

To build a good simulation model of the Google cluster, we needed to extract some information from the traces. The data consist of a large amount of CSV files containing records about job and task events, resources used by tasks, task constraints, etc. There are more than 2000 files describing the workload and machine attributes for over 12000 cluster nodes, reaching a total \emph{compressed size} of about 40~GB. In total, over 1.3 billion records are available. We used BiDAl to extract the arrival time distribution of each job, the distribution of the number of tasks per job, and the distributions of execution times of different types of tasks (e.g., jobs that successfully completed execution, jobs that are killed by the users, and so on). These distributions are used by the Job Arrival entity of the simulation model to generate jobs into the system. Additionally, we analyzed the distribution of machines downtime and of the time instants when servers are added / removed from the pool. 

Some of the results obtained with BiDAl are shown in the following (we are showing the actual plots that are produced by our software). Figure~\ref{ram} shows the the amount of RAM requested by tasks, while Figure~\ref{task} shows the distribution of number of tasks per job.

To generate the graph in Figure~\ref{task}, we first extracted the relevant information from the trace files.  Job and task IDs were required, therefore we generated a new table, called \emph{job\_task\_id}, from the \emph{task\_events.csv} files released by Google \cite{googleData}. The query generation is automated by BiDAl which allows for simple selection of columns using the GUI. Since the DISTINCT clause is not yet implemented in BiDAl, we added it manually in the generated query.  The final query used was:

\vspace{-3mm}
{\footnotesize \begin{verbatim}SELECT DISTINCT V3 AS V1,V4 AS V2 FROM task_events \end{verbatim}
}
\vspace{-3mm}

where V3 is the \emph{job\_id} column while V4 represents the \emph{task\_id}.On the resulting \emph{job\_task\_id table}, we execute another query to estimate how many tasks each job has, generating a new table called \emph{tasks\_per\_job}:

\vspace{-3mm}
{\footnotesize\begin{verbatim}SELECT V1 AS V1, COUNT(V2) AS V2 FROM job_task_id GROUP BY V1\end{verbatim}
}
\vspace{-3mm}

Three R commands were used on the \emph{tasks\_per\_job} table to generate the graph. The first extracts the second column (job id), the second filters out some uninteresting data and the third plots the result. The BiDAl commands used are shown in Table~\ref{ex}.

The analysis was performed on a computer with 16~GB of RAM, a 2.7 GHz i7 quad core processor and a hard drive with simultaneous read/write speed of 60~MB/s. For the example above, importing the data was the most time consuming step, requiring 11 minutes to load 17~GB of data into the SQLite storage (which may be influenced by the disk speed). However, this step is required only once. The first SQL query took about 4 minutes to complete, while the second query and the R commands were almost instantaneous.

\begin{figure}
  \begin{center}
    \includegraphics[width=0.87\textwidth]{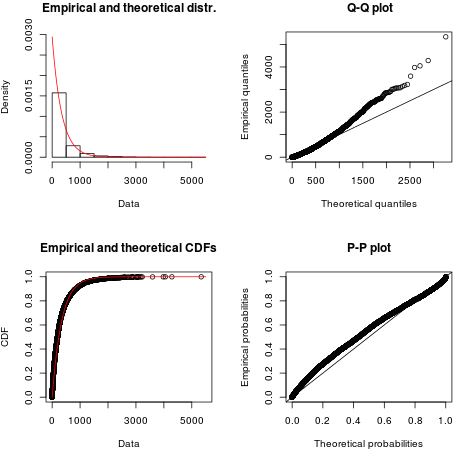}
    \caption{\label{exp}Machine update events, fitted with  an exponential distribution. The left panels show the density and cumulative distribution functions, with the lines representing the exponential fitting and the bars/circles showing real data. The right panels show goodness of fit in Q-Q and P-P plots (straight lines show perfect fit).}
  \end{center}
\end{figure}

In Figure~\ref{exp} we tried to fit the time between consecutive machine update events (i.e., events that indicate that a machine has changed its list of resources) with an exponential distribution; the four standard plots for the goodness of fit show that the observed data is in good agreement with the fitted distribution.

\begin{figure}
  \begin{center}
    \begin{subfigure}{0.45\textwidth}
    \includegraphics[width=\textwidth]{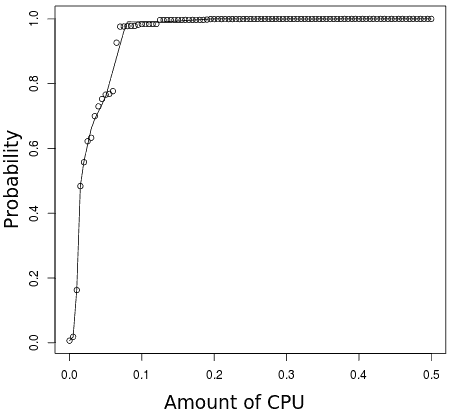}\caption{CPU task requirements}\label{cpuCdf}
    \end{subfigure}
     \begin{subfigure}{0.45\textwidth}
    \includegraphics[width=\textwidth]{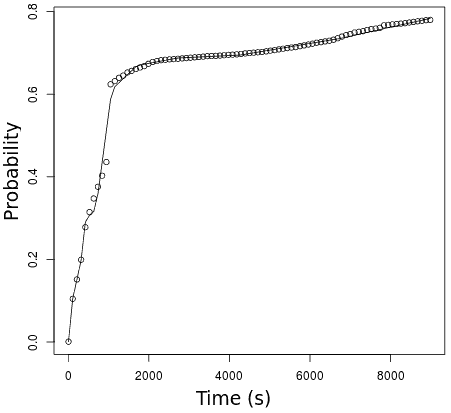}\caption{Machine downtime}\label{downtimeCdf}
    \end{subfigure}
    \caption{Examples of CDFs fitted by sequences of splines, obtained with BiDAl. The circles represent the data, while the lines show the fitted splines.}
  \end{center}
\end{figure}

Cumulative distribution functions (CDFs) have also been obtained from the data and fitted with sequences of splines, in those cases where the density functions were too noisy to be fitted with a known distribution. For instance, Figure~\ref{cpuCdf} shows the distribution of CPU required by tasks while Figure~\ref{downtimeCdf} shows machine downtime, both generated with BiDAl. Several other distributions were generated, similar to CPU requirements, to enable simulation of the Google cluster:
 RAM required by tasks; 
 Tasks priority;
 Duration of tasks that end normally;
 Duration of killed tasks;
 Tasks per job;
 Job inter-arrival time;
 Machine failure inter-arrival time;
 Machine CPU and RAM.

\subsection{Cluster Simulator}

We built a discrete-event simulation model of the Google compute cluster corresponding to that from which the traces were obtained, using C++ and Omnet++. According to the information available, the Google cluster is basically a large batch system where computational tasks of different types are submitted and executed on a large server pool. Each job may describe constraints for its execution (e.g., a minimum amount of available RAM on the execution host); a scheduler is responsible for extracting jobs from the waiting queue, and dispatching them to a suitable execution host. As can be expected on a large infrastructure, jobs may fail and be resubmitted; moreover, execution hosts may fail and be temporarily removed from the pool, or new hosts can be added. The Google trace contains a list of timestamped events such as job arrival, job completion, activation of a new host and so on; additional (anonymized)  information on job requirements is also provided.

\begin{figure}
  \begin{center}
    \includegraphics[width=0.7\textwidth]{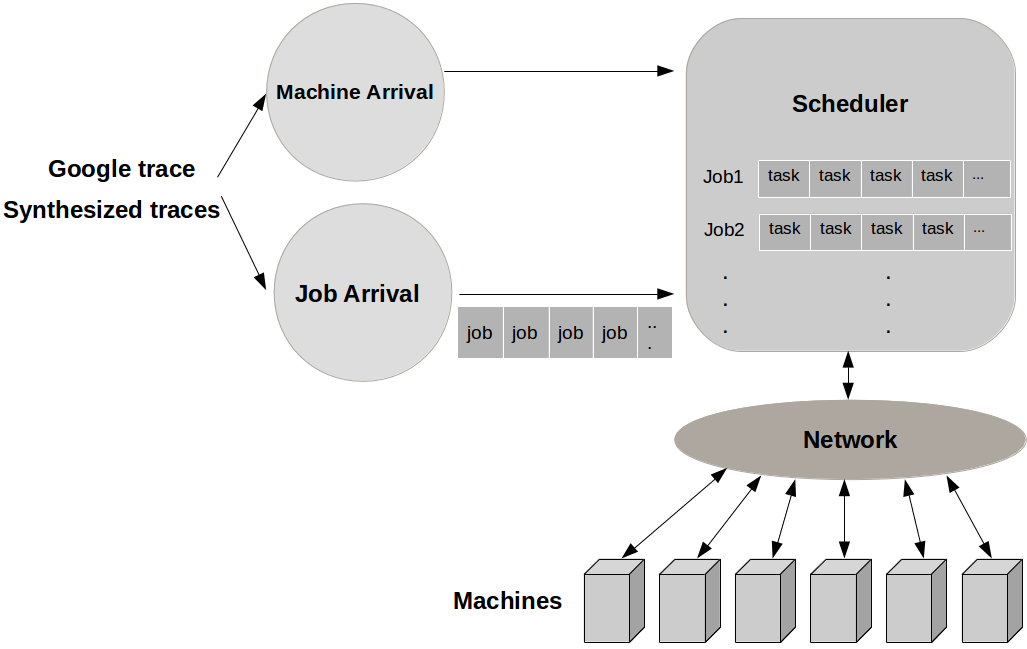}
    \caption{\label{sim}Architecture of simulation model.}
  \end{center}
\end{figure}

The simulation model, shown in Figure~\ref{sim}, consists of several active and passive interacting entities. The passive entities (i.e., those that do exchange any message with other entities) are Jobs and Tasks. A task represents a process in execution, or ready to be executed; each task has an unique ID and the amount of resources required; a Job is a set of (likely dependent) tasks. Jobs can terminate either because all their tasks complete execution, or because they are aborted by the submitting user.

The active entities in the simulation are those that  send and receive messages: Machine, Machine Arrival, Job Arrival, Scheduler and Network. The Machine entity represents an execution node in the compute cluster. Machine Arrival and Job Arrival generate events related to new execution nodes being added to the cluster, and new jobs being submitted, respectively. These arrival processes (as they are called in queueing theory) can be driven by the real trace logs, or by synthetic data generated from user-defined probability distributions that can be identified using BiDAl. The Scheduler implements a simple job scheduling mechanism. Every time a job is created by the JobArrival entity, the scheduler inserts its tasks in the waiting queue. For each task, the scheduler examines which execution nodes (if any) match the task constraints; the task is eventually sent to a suitable execution node. Note that the scheduling policies implemented by the Google cluster allow a task with higher priority to evict an already running task with lower priority; this eviction priority mechanism is also implemented in our simulator. Finally, the Network entity is responsible for simulating the message exchanges between the other active entities.

\subsection{Trace-Driven Simulation of the Google Cluster}

We used the parameters extracted from the traces to instantiate and run the simulation model. From the the traces, it appeared that the average memory usage of the Google machines is more or less constant at~50\%. According to Google, the remaining memory on each server is reserved to internal processes. Therefore, in the simulation we also set the maximum available memory on each server at half the actual amount of installed RAM.

\begin{figure}
  \begin{center}
    \begin{subfigure}{0.49\textwidth}
    \includegraphics[width=\textwidth]{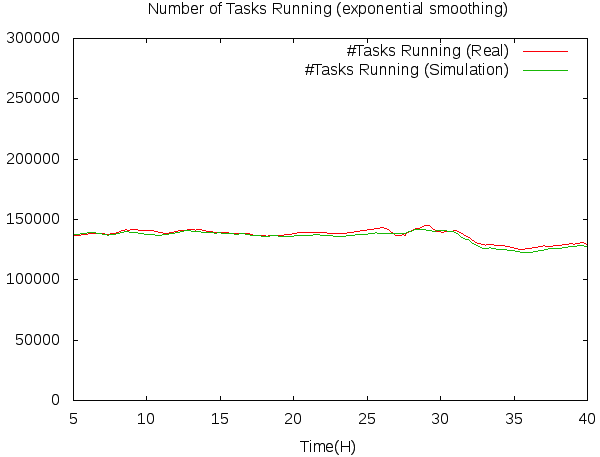}\caption{Number of running tasks.}\label{running}
    \end{subfigure}
     \begin{subfigure}{0.49\textwidth}
    \includegraphics[width=\textwidth]{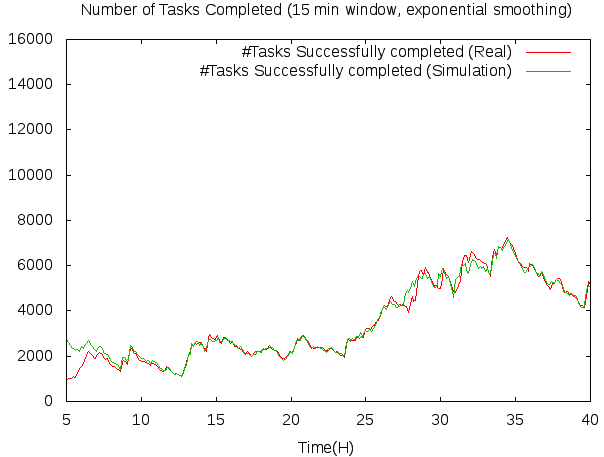}\caption{Number of tasks completed.}\label{completed}
    \end{subfigure}
     \begin{subfigure}{0.49\textwidth}
    \includegraphics[width=\textwidth]{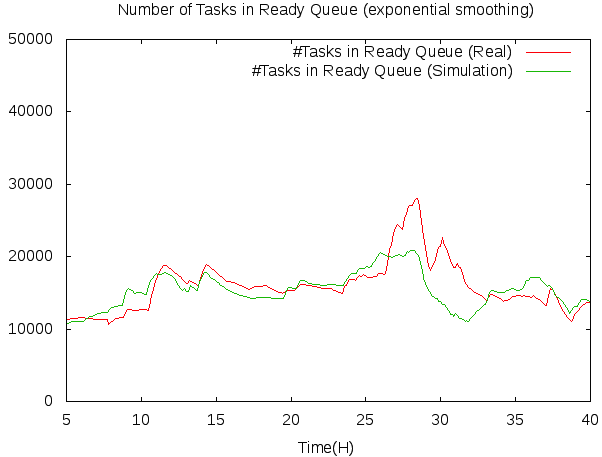}\caption{Number of tasks waiting.}\label{waiting}
    \end{subfigure}
     \begin{subfigure}{0.49\textwidth}
    \includegraphics[width=\textwidth]{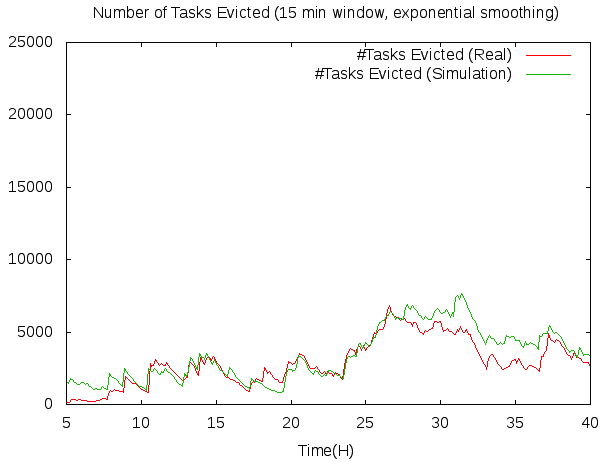}\caption{Number of tasks evicted.}\label{evicted}
    \end{subfigure}
    \caption{Simulation and real data for four different metrics. }
  \end{center}
\end{figure}

The purpose of the simulation run was to validate the model by comparing the real traces with simulator results. Four metrics were considered: number of running tasks (Figure~\ref{running}), completed tasks (Figure~\ref{completed}), waiting tasks (ready queue size, Figure~\ref{waiting}) and evicted tasks (Figure~\ref{evicted}). All plots show the time series extracted from the trace data (green lines) and those produced by our simulator (red lines), with the additional application of exponential smoothing to reduce transient fluctuations. The figures show a very good agreement between the simulation results and the actual data from the traces.

\section{Related work}\label{refs}

With the public availability of the two Google cluster traces~\cite{googleData}, numerous analyses of different aspects of the data have been reported. These provide general statistics about the workload and node state for such clusters~\cite{Liu2012,Reiss2012,Reiss2012a} and identify high levels of heterogeneity and dynamicity of the system, especially in comparison to grid workloads~\cite{Di2012}. However, no unified tool for studying the different traces were introduced. BiDAl is one of the first such tools facilitating Big Data analysis of trace data, which underlines similar properties of the public Google traces as the previous studies. Other traces have been analyzed in the past~\cite{Kavulya2010,Chen2011,Chen2012}, but again without a dedicated tool available for further study.

BiDAl can be very useful in generating synthetic trace data. In general synthesising traces involves two phases: characterising the process by analyzing historical data and generation of new data. The aforementioned Google traces and log data from other sources have been successfully used for workload characterisation. In terms of resource usage, classes of jobs and their prevalence can be used to characterize workloads and generate new ones~\cite{Mishra2010,Wang2011}, or real usage patterns can be replaced by the average utilization~\cite{Zhang2011}.  Placement constraints have also been synthesized using clustering for characterisation~\cite{Sharma2011}. Our tool enables workload and cloud structure characterisation through fitting of distributions that can be further used for trace synthesis. The analysis is not restricted to one particular aspect, but the flexibility of our tool allows the the user to decide what phenomenon to characterize and then simulate.

Recently, the Failure Trace Archive (FTA) has published a toolkit for analysis of failure trace data~\cite{Javadi2013}. This toolkit is implemented in Matlab and enables analysis of traces from the FTA repository, which consists of about 20 public traces. It is, to our knowledge, the only other tool for large scale trace data analysis. However, the analysis is only possible if traces are stored in the FTA format in a relational database, and is only available for traces containing failure information. BiDAl on the other hand provides two different storage options, including HDFS, with transfer among them transparent to the user, and is available for any trace data, regardless of what process it describes. Additionally, usage of FTA on new data requires publication of the data in their repository, while BiDAl can be used also for sensitive data that cannot be made public. 

Although public tools for analysis of general trace data are scarce, several large corporations have reported building in-house applications for analysis of logs. These are, in general, used for live monitoring of the system, and analyze in real time large amounts of data to provide visualisation that helps operators make administrative decisions. While Facebook use Scuba~\cite{Abraham2013}, mentioned before, Microsoft have developed the Autopilot system~\cite{Isard2007}, which helps administer their clusters. This has a component (Cockpit) that analyzes logs and provides real time statistics to operators. An example from Google is CPI2~\cite{Hagmann2013} which monitors Cycles per Instruction for running tasks to determine job performance interference. This helps in deciding task migration or throttling to maintain high performance of production jobs. All these tools are, however, not open, apply only to data of the corresponding company and sometimes require very large computational resources (e.g. Scuba). Our aim in this paper is to provide an open research tool that can be used also by smaller research groups that have more limited resources.

\section{Conclusions}\label{conclusions}
In this paper we presented BiDAl, a framework that facilitates use of Big Data tools and techniques for analyzing large cluster traces. BiDAl is based on a modular architecture, and currently supports two storage backends based on SQlite and Hadoop; other backends can be easily added. BiDAl uses a subset of SQL as a common query language that is automatically translated to the appropriate commands supported by each backend. Additionally, data analysis using R and Hadoop MapReduce is possible. 

We have described a usage example of BiDAl that involved the analysis of Google trace data to derive parameters to be used in a simulation model of the Google cluster. Distributions of relevant quantities were easily computed using our tool, showing how this facilitates Big Data analysis even to users less familiar with R or Hadoop. Using the computed distributions, the simulator produces results that are in good agreement with the observed data. Another possible usage of the platform is for application of machine learning tools for predicting abnormal behavior from log data. At the moment, BiDAl can be used for pre-processing and initial data exploration; however, in the future we plan to add new commands to perform this type of analysis directly. Both usage examples could provide new steps towards achieving self-* properties for large scale computing infrastructures in the spirit of Autonomic Computing.
In its current implementation, BiDAl is useful for batch analysis of historical log data, which is important for modeling and initial training of machine learning algorithms. However, live log data analysis is also of interest, so we are investigating the addition of an interface to streaming data sources to our platform. Future work also includes implementation of other storage systems, especially to include non-relational models. Improvement of the GUI and general user experience will also be pursued.

\end{document}